\documentclass[10pt,letterpaper]{article}
\usepackage{opex3}
\usepackage{color}
\usepackage{comment}
\usepackage{tikz}
\usepackage{pgfplots}
\usepackage{textcomp}
\usepackage{color,soul}
\usepackage[english]{babel}

\usepgfplotslibrary{groupplots}

\pgfplotsset{compat=newest}

\graphicspath{{./images/}}
\begin{document}

\title{Analysis of 3D-printed metal for rapid-prototyped reflective terahertz optics}

\author{Daniel Headland,$^{1,*}$ Withawat Withayachumnankul,$^{1,\dagger}$ Michael Webb,$^{2}$ Heike Ebendorff-Heidepriem,$^{3}$ Andre Luiten$^{3}$ and Derek Abbott$^{1}$}

\address{$^1$School of Electrical and Electronic Engineering, The University of Adelaide, \\ SA 5005, Australia\\
$^2$Centre for Defence Communications and Information Networking, The University of Adelaide, SA 5005, Australia\\
$^3$Institute for Photonics and Advanced Sensing, The University of Adelaide, SA 5005, Australia}

\email{$^*$daniel.headland@adelaide.edu.au} 
\email{$^\dagger$withawat.withayachumnankul@adelaide.edu.au} 



\begin{abstract}
We explore the potential of 3D metal printing to realize complex conductive terahertz devices. 
Factors impacting performance such as printing resolution, surface roughness, oxidation, and material loss are investigated via analytical, numerical, and experimental approaches. 
The high degree of control offered by a 3D-printed topology is exploited to realize a zone plate operating at 530 GHz. 
Reflection efficiency at this frequency is found to be over 90\%.
The high-performance of this preliminary device suggest that  3D metal printing can play a strong role in guided-wave and general beam control devices in the terahertz range.
\end{abstract}







\section{Introduction}

In recent years, 3D-printing, a form of additive manufacture, has gained significant attention, in part due to its generality, versatility, and compatibility with computer-aided design tools. 
It is well-suited to rapid prototyping, small- to medium-scale manufacture, and fabricating replacement parts and custom equipment on demand \cite{berman20123,anzalone2013low,zhang2013open}.
Additionally, 3D-printing technology has been demonstrated in cutting-edge applications to construct compact microbatteries \cite{sun20133d}, miniaturized chemical reactors \cite{kitson2012configurable,symes2012integrated}, scaffolding for visceral \cite{kim1998survival} and bone tissue growth \cite{seitz2005three}, and custom surgical implants and models \cite{seitz2005three,habibovic2008osteoconduction}.

The ability of 3D printing to realize complex and high-precision structures opens a path for printing devices that manipulate electromagnetic radiation.
For example, 3D-printed lenses that operate in the optical range are presently a commercially available product \cite{blomaard20153d}. 
In the terahertz range, there is a need for rapid prototyping techniques to accelerate the development of practical technologies. 
The resolution of 3D-printers is typically in the order of a few tens to a few hundreds of micrometers.
Such a scale is highly suitable for the manipulation of terahertz waves, with a wavelength spanning from 30~\textmu{}m to 1~mm, as complicated topologies can readily be realized at a scale comparable to a wavelength. 
Furthermore, the achievable dimensions of 3D-printed structures are in the order of several hundreds of wavelengths, making it possible to realize devices of large aperture. 
As such, 3D-printing of polymer dielectrics has recently been employed to realize numerous devices for shaping terahertz radiation, including conventional lenses \cite{busch2014optical,busch2015thz,suszek20153,guo20152,squires20153d}, phase plates \cite{zhu2014experimental,wei2014orbit}, axicons \cite{liu2014terahertz}, and reflectarrays \cite{nayeri20143d}. 
On the other hand, metals are naturally suited to reflectives devices and guiding structures, which are critical devices in the terahertz range. 
Techniques such as selective laser melting (SLM) \cite{thijs2010study,bremen2012selective} make it possible to 3D-print directly in solid metals  \cite{ladd20133d,frazier2014metal}.
Despite this, there have been no demonstrations of direct metal printing in the context of terahertz technologies. 
The most related work is terahertz guided-wave structures that were made from 3D-printing of polymers, with a metal layer deposited on the surface  \cite{pandey2013terahertz,liu2013maxwell}. 
The application of direct 3D-printing of metals offers an important nice for rapid and versatile realization of such devices. 

In this work, we investigate the previously unexplored approach of using direct 3D-printing of metals to realize terahertz devices. 
We select grade-5 titanium for the build material, as it is a common titanium alloy of wide-spread use in areas including aerospace, automotive, marine, and medical applications \cite{welsch1993materials}.
The efficiency of this alloy as a reflector is investigated using experimental and analytical means. 
As a demonstration, a 3D-printed metal phased zone plate operating at 530~GHz is realized, and its capacity to focus terahertz radiation is demonstrated. 
The operating frequency of 530~GHz is selected as suitable to achieve a binary phase difference, accommodating a conservative estimate of the vertical resolution limitations imposed by the 3D-printing procedure. 

\begin{figure}[b]
\centering
\includegraphics{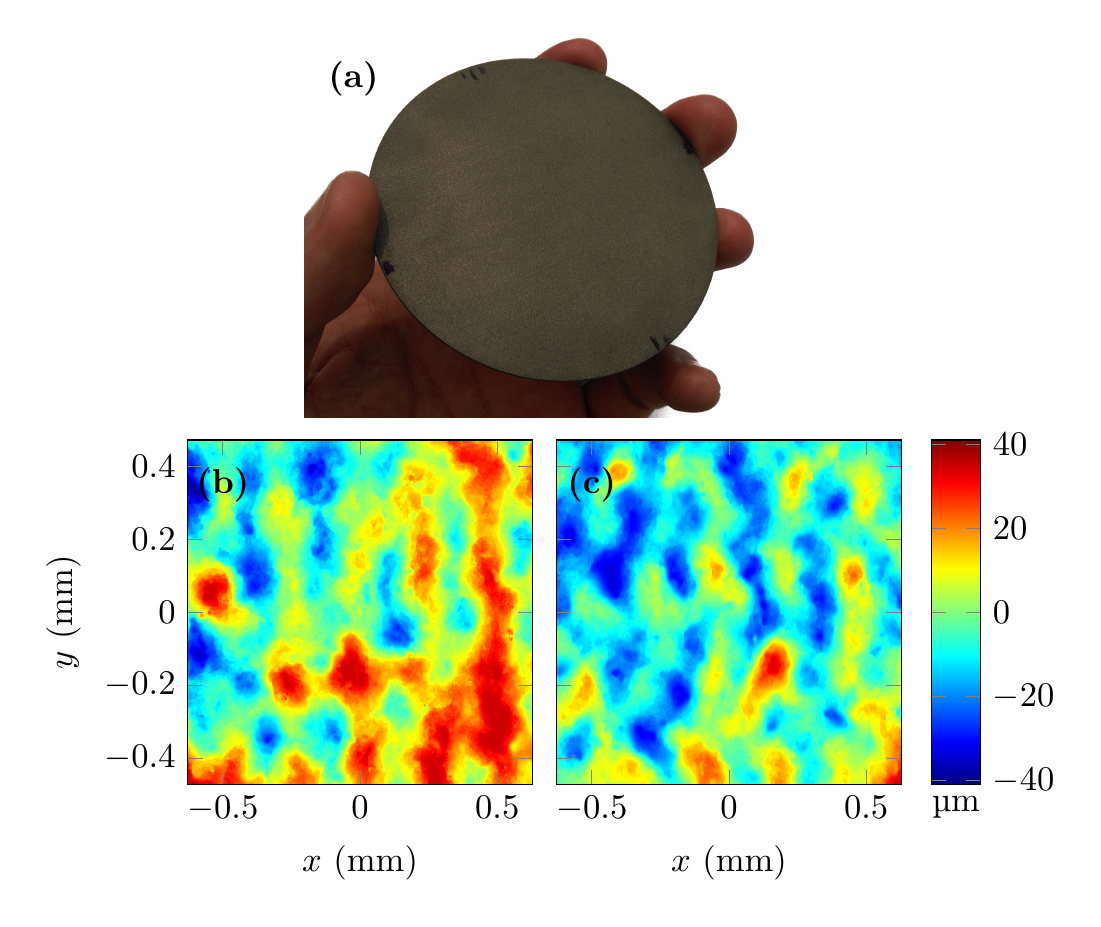}
%
\caption{Fabricated flat metal disk, showing (a) photograph of $\sim$75~mm-diameter sample, and (b,c) optical profiler data at two different locations on the sample surface, with standard deviations $\sigma_h$ of 10.78 and 10.28~\textmu{}m respectively }
\label{fig:fabsample}
\end{figure}

\section{Characteristics of 3D-printed metal}

\subsection{Fabrication}

In the 3D-printing technique known as SLM, a structure is built-up in a layered fashion, by alternate passes of deposition of the  powder build material, and selective exposure to a high-intensity raster-scanning laser. 
The laser melts a certain portion of the powder material, thereby fusing, and subsequently cooling down, to form the solid portions of the relevant layer \cite{bremen2012selective}. 
The present work makes use of a ProX 200 SLM printer, which employs a 1070-nm laser at 300~W. 
The spot size of the laser is approximately 70~\textmu{}m, and the scanning speed is 1,800~mm/sec. 
The printing procedure takes place in an argon gas chamber, with oxygen averaging at 1000~ppm. 
The metal powder utilized is a titanium alloy that is commonly known as ``Grade-5 titanium", ``Ti6Al4V", ``Ti6-Al4-V", or ``Ti 6-4".
It is composed of $\sim$90\% titanium, $\sim$6\% aluminum, and $\sim$4\% vanadium.
The particle size of the metal powder employed in 3D-printing is at most 40~\textmu{}m. 
Post 3D-printing fabrication, the structure is annealed in an open-air furnace at 650$^\circ$ for a period of two hours in order to relieve stresses that are created within the material during the build process. 
Annealing results in the development of an oxidation layer, as an un-intended side-effect. 
Finally, the sample is sandblasted at 600~kPa with a particle mixture that is 70\% garnet and 30\% glass, and particle sizes in the range from 90 to 400~\textmu{}m, in order to smooth the surface. 

For characterization purposes, a flat, featureless metal disk is fabricated, which is $\sim$75~mm in diameter and $\sim$1~mm thick. 
This sample is shown in Fig.~\ref{fig:fabsample}(a).
Its surface roughness is determined using an optical interferometic profiler (model Contour GT–K1 Optical Profiler Stitching System from Veeco), and acquired images are presented in Fig.~\ref{fig:fabsample}(b,c).
There appears to be some quasi-periodic regularity to the roughness in the form of grains of several hundred micrometers. 
This is a consequence of the dynamics of the liquid-phase metal during the fabrication procedure, combined with the sequential hatching of the laser \cite{thijs2010study}.
As shown in Fig.~\ref{fig:fabsample}(b,c), optical profiler measurements of the 3D-printed metal reveal the standard deviation of the surface height to be $\sigma_h=10$--$11$~\textmu{}m, which is in the order of the particle size employed in 3D-printing. 
Note the value $\sigma_h$ can alternatively be called the root mean squared of the surface perturbation, commonly denoted $S_q$, as the mathematical definitions are identical. 

In order to determine the presence of oxygen, and any possible contaminations in the surface layer after annealing and sandblasting, energy-dispersive X-ray spectroscopy (EDX) is performed. 
Four distinct compounds are identified in the surface layer.
Most representative in the surface is titania, followed by alumina. 
This is to be expected, as titanium and aluminium readily oxidise in air at the annealing temperature of 650$^\circ$. 
There are also pieces of garnet embedded in the surface, as a result of the sandblasting steps. 
Finally, there are particles of carbon-rich material, most likely dust, on the surface. 
Note, the presence of vanadium was not detected in EDX characterization, due to low concentration in the alloy used. 

\subsection{Experiment}

\begin{figure}[tb]
\centering
\includegraphics{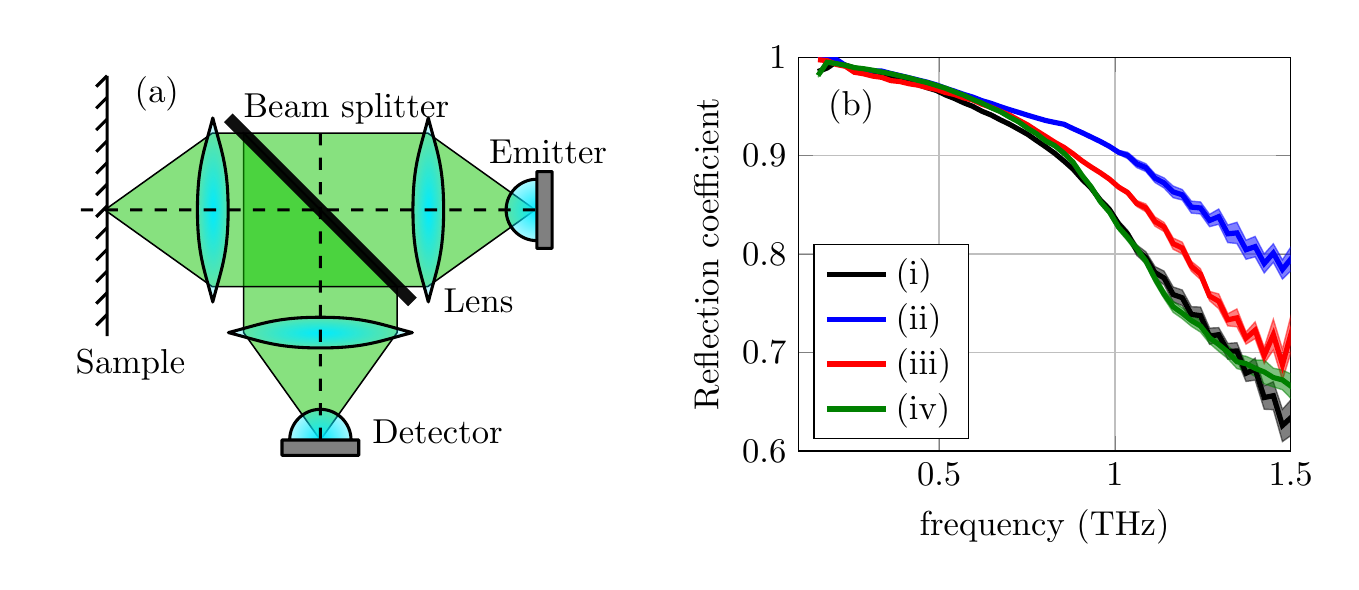}
\caption{Characterization of 3D-printed titanium alloy (a) measurement setup (b) measured reflection coefficient at normal incidence, with error bars shown as colored regions. (i-iv) Different locations on the sample surface.}
\label{fig:reflectivity_setup}
\end{figure}

In order to determine the efficiency of devices created with  3D-printed grade-5 titanium, it is necessary to characterize the material for its terahertz properties. 
The reflectance of the flat sample given in Fig.~\ref{fig:fabsample}(a) is characterized with terahertz time-domain spectroscopy (THz-TDS), using the normally-incident reflection setup given in Fig.~\ref{fig:reflectivity_setup}(a). 
A focused beam is used to probe the properties of the metal, in order to effectively isolate the surface properties at a given point, and determine the degree to which these properties vary across the surface. 
The beam waist of the focused beam is $\sim$1~mm, which results in a Rayleigh range of $\sim$5~mm. 
As the terahertz radiation is incident on the sample at the approximate location of the beam waist, normal incidence of all rays can be assumed. 

Measurements are taken at four arbitrary points on the sample surface.
Five measurements are taken for each point, and the averaged results are given in Fig.~\ref{fig:reflectivity_setup}(b), with standard deviations given as colored regions. 
There are significant discrepancies between the different sets of measurements at frequencies above $\sim$700~GHz, which suggests a variation in reflectance across the sample surface. 
The efficiency of a reflector of this sort is defined as the square of the reflection coefficient magnitudeAround the nominal operating frequency of 530~GHz, the reflection coefficient is over 95\%, which is equivalent to efficiency of over 90\%. 
In all cases, for frequencies below 900~GHz, overall efficiency is greater than 80\%. 

\subsection{Modeling of reflection characteristics}

\label{reflection}

A model is developed in order to explain the reflection response of the 3D-printed metal. 
This model takes into account losses due to both material dissipation and surface irregularities. 
The reflection coefficient, $\rho_\mathrm{r}$, considering only scattering loss due to surface roughness can be determined using Eq.~\ref{eq:roughness} \cite{dikmelik2006roughness}, where $\sigma_h$ represents the standard deviation of the surface height
\begin{equation}
\label{eq:roughness}
\rho_\mathrm{r} = \mathrm{exp} \left[ - 2 \left(\frac{\omega \sigma_h}{c}\right)^2 \right].
\end{equation}
This expression is based on the summation of delayed responses introduced by perturbations in the $z$-position of the surface, and is independent of polarization. 
The result is effectively a form of low-pass filter, with Gaussian roll-off.

\begin{figure}[tb]
\centering
\includegraphics{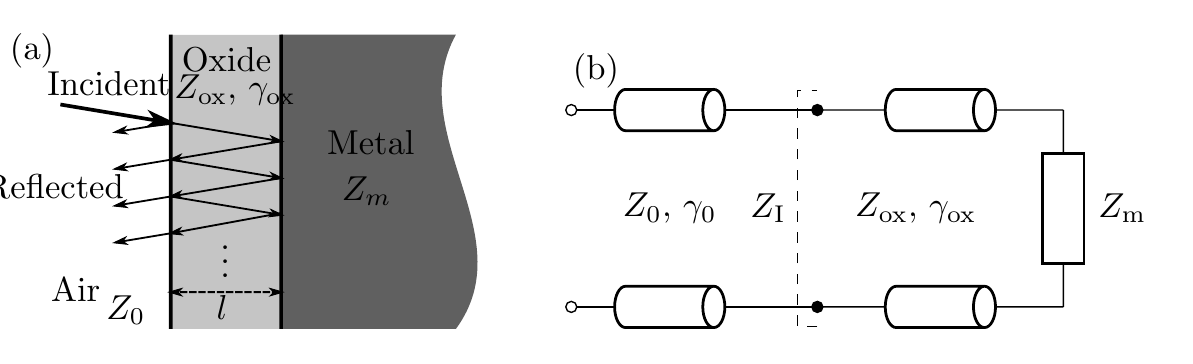}

\caption{ Reflection of radiation from sample surface, neglecting surface roughness, showing 
(a) oxide-on-metal structure, with internal reflection in the dielectric layer, and
(b) Equivalent transmission-line model, where $Z_\mathrm{I}$ is the input impedance of the surface. }
\label{fig:TLM}
\end{figure}

\begin{figure}[t]
\centering
\includegraphics{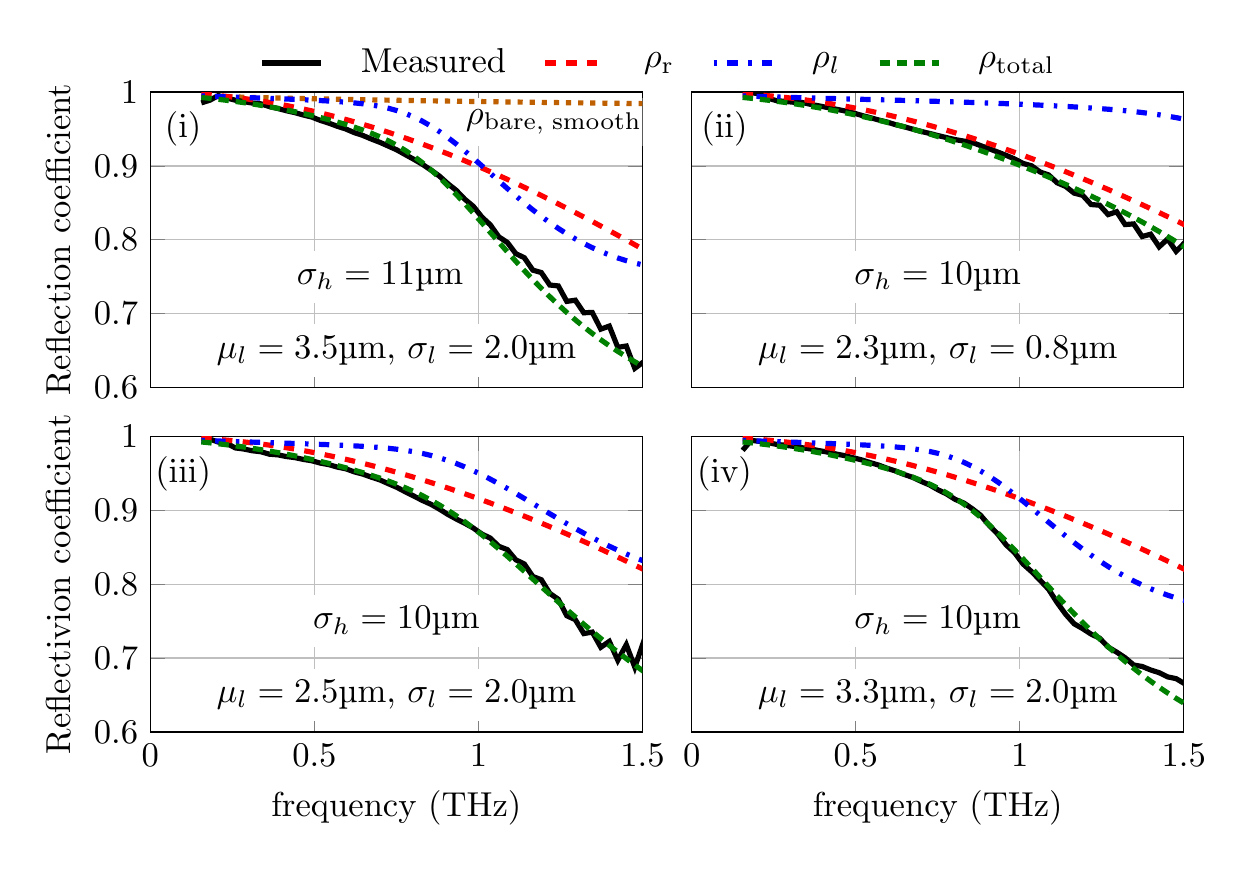}
\caption{Modeling of printed metal, fit to measured results, where (i-iv) correspond to measurements in Fig.~\ref{fig:reflectivity_setup}(b). 
Scattering reflection coefficient, transmission line model reflection coefficient, and overall modeled reflection coefficient are given by  $\rho_\mathrm{r}$, $\rho_l$, and  $\rho_\mathrm{total}$, respectively.
The fitting parameters $\sigma_h$, $\mu_l$, and $\sigma_l$ are the standard deviation of surface roughness, and the mean and standard deviation of oxide layer thickness, respectively.
The reflection coefficient of perfectly smooth grade-5 titanium metal is also given in (i), as $\rho_{\mathrm{bare,\; smooth}}$.
 }
\label{fig:reflectivity_model}
\end{figure}

Another source of loss that must be considered is dissipation, both in the bulk metal, and in a thin oxidation layer on the surface.
A transmission-line model, illustrated in Fig.~\ref{fig:TLM}, is employed for this system.
The impedance, $Z$, and propagation constant, $\gamma$, of a given material are calculated using Eqs.~\ref{eq:impedance},\ref{eq:propconst} \cite{pozar2009microwave}, where $\mu_0$ and $\epsilon_0$ are the vacuum magnetic permeability and electric permittivity respectively, and $\epsilon_r - j \epsilon_i$ is the complex relative permittivity of the relevant medium, 
\begin{equation}
\label{eq:impedance}
Z = \sqrt{ \frac{\mu_0}{\epsilon_0 (\epsilon_r - j \epsilon_i)}},
\end{equation}
\begin{equation}
\label{eq:propconst}
\gamma = j \omega \sqrt{ \mu_0 \epsilon_0 (\epsilon_r - j \epsilon_i)}.
\end{equation}
The wave impedances for metal and oxide are given by $Z_\mathrm{m}$ and $Z_{\mathrm{ox}}$ respectively, and the complex propagation constant in the oxide is given by $\gamma_{\mathrm{ox}}$. 
The material properties of grade-5 titanium are determined in accordance with a Drude model \cite{wieting1976infrared,welsch1993materials}, as elaborated in Appendix~\ref{materialdrude}.
Grade-5 titanium is 90\% titanium, and hence titania is expected to make up the majority of the oxide layer, as has been confirmed by EDX characterisation. 
Furthermore, although there are impurities in the form of alumina, garnet, and dust, the relative permittivity of titania is extremely high ($\epsilon_r \sim 110$, see Appendix~\ref{materialoxide}), and hence titania dominates the response of the material.  
Therefore, the oxide layer is approximated as pure titania, in order to simplify analysis. 
Lower-index impurities such as alumina will simply result in a local reduction in the effective thickness of the oxide layer. 
As the thickness of the oxide layer is not precisely known, it is treated as a normally-distributed random variable $L$, with mean and standard deviation $\mu_l$ and $\sigma_l$ respectively. 
The input impedance of the transmission line system,  $Z_\mathrm{I}$, and hence the reflection coefficient, $\rho_\mathrm{TLM}$, are therefore random variables as well, as follows
\begin{equation}
\label{eq:inputZ}
Z_\mathrm{I}(L) = Z_{\mathrm{ox}} \frac{Z_\mathrm{m} + Z_{\mathrm{ox}} \mathrm{tanh}(\gamma_{\mathrm{ox}} L)}{Z_{\mathrm{ox}} + Z_\mathrm{m} \mathrm{tanh}(\gamma_{\mathrm{ox}} L)},
\end{equation}
\begin{equation}
\label{eq:Gamma}
\rho_\mathrm{TLM}(L) = \frac{Z_\mathrm{I}(L) - Z_0}{Z_\mathrm{I}(L) + Z_0}.
\end{equation}
Given that an incident beam will cover an area of the surface, the overall reflected beam will effectively average the distribution of oxide thicknesses. 
Hence the reflection coefficient of the system according to the transmission-line model, $\rho_l$, is the expected value of the reflection coefficient of the transmission line system $\rho_\mathrm{TLM}$, or
\begin{equation}
\label{eq:rt}
\rho_l = \langle \rho_\mathrm{TLM}(L) \rangle = \frac{\langle Z_\mathrm{I}(L) \rangle - Z_0}{ \langle Z_\mathrm{I}(L) \rangle + Z_0}.
\end{equation}
The expected value of the input impedance can be computed with the following integral, which discounts non-physical negative values of the thickness,
\begin{equation}
\label{eq:integral}
\langle Z_\mathrm{I}(L) \rangle = \frac{1}{\sigma_l \sqrt{2 \pi}}\int_0^\infty Z_\mathrm{I}(l) \mathrm{exp} \left( -\frac{(l-\mu_l)^2}{2 \sigma_l^2} \right) \mathrm{d} l.
\end{equation}
This integral is computed numerically.
Note that, in order for the probability density function of the truncated Gaussian distribution to be valid, the result in Eq.~\ref{eq:integral} must be normalized by the factor $1 - I_{\mathrm{lower}}$, where 
\begin{equation}
\label{eq:lowerbound}
I_{\mathrm{lower}} = \frac{1}{\sigma_l \sqrt{2 \pi}}\int_{-\infty}^0 \mathrm{exp} \left( -\frac{(t-\mu_l)^2}{2 \sigma_l^2} \right) \mathrm{d} l.
\end{equation}
This is to ensure that the total integral of the probability density function sums to 1. 
Although this means the value $\mu_l$ is no longer the true mean of the distribution, it does not correspond to any directly measured value, and hence this is of no consequence. 
The overall reflection coefficient is computed by taking the product of the reflectivities due to the surface roughness and the loss, or
\begin{equation}
\label{eq:total}
\rho_\mathrm{total} = \rho_\mathrm{r}  \rho_l.
\end{equation}
The statistical properties of the roughness and the oxidation layer thickness are treated as free parameters, in order to match them to each set of measured results.
Results of this procedure are presented alongside measured results in Fig.~\ref{fig:reflectivity_model}.
For further insight, the theoretical reflection coefficient of perfectly smooth, bare grade-5 titanium is determined, and is presented in Fig.~\ref{fig:reflectivity_model}(i). 
This is determined by substituting $Z_\mathrm{m}$ for $Z_\mathrm{I}$ in Eq.~\ref{eq:Gamma}.

Strong agreement is achieved between the measured results and the model. 
Additionally, the standard deviation of surface level is in agreement with the optical profiler measurements presented in Fig.~\ref{fig:fabsample}.
It can be seen that the statistical properties of the surface topology and oxide layer thickness play a significant role in the frequency-dependent response. 
Furthermore, given that the measured response is different for measurements taken at different points on the sample surface, these statistical properties must vary across the surface of the sample. 
Lastly, reflection coefficients $\rho_l$ and $\rho_r$ are significantly lower than the reflection coefficient of idealized bare grade-5 titanium, and hence we assert that surface roughness and the oxide layer are the most significant contributors to reflection loss, especially at higher frequencies.  

\section{3D-printed zone plate}

A binary phased zone plate consists of alternate rings of 0 and $\pi$ phases, in a concentric arrangement, in order to focus radiation via diffraction. 
Whilst such a structure could potentially have been fabricated by machining, our work serves as proof-of-concept for arbitrary 3D-printed reflective devices such as binary-phase holorams \cite{yang2014fast}, which are more difficult to directly machine.
A zone plate has been chosen for this purpose, as the performance can be characterized in a relatively straightforward manner.

\subsection{Required phase distribution}

Beam control is most commonly achieved by imposing a particular phase profile on a wavefront, which determines the progression of the beam. 
Whilst reflectarrays and phased arrays achieve the bespoke phase profile by resonators or electronic phase shifters \cite{niu2013experimental,zou2013dielectric,niu2014terahertz,headland2015terahertz}, in the case of a 3D-printed device it is achieved by path difference in the topology itself. 
As the vertical resolution of additive manufacture is limited, the minimum achievable path difference, and hence phase difference, is quantized into levels. 
If the radiation is obliquely incident on the structure, with an angle of incidence of $\theta$, the minimal phase shift is dictated by the following equation, where the vertical resolution is $\Delta h$
\begin{equation}
	\Delta \phi = \frac{4 \pi}{\lambda} \Delta h \; \mathrm{cos} \theta.
	\label{eq:oblpath}
\end{equation}
Note that a larger angle of incidence will result in finer phase quantization. 

Given that the vertical resolution is fixed, the maximum operating frequency is dictated wholly by the choice of phase quantization. 
For a given vertical resolution, a larger phase quantization value will result in a higher operating frequency. 
The maximal viable phase quantization is $\pi$ radians, and this results in binary phase, which is sufficient for diverse beam-shaping applications \cite{rastani1991binary,dufresne2001computer,ginn2008planar,yang2014fast,zhang2015three}.
In this case, the conservative vertical resolution of the 3D printer is 200~\textmu{}m, and hence the operating frequency of binary optics is 375~GHz for normal incidence, and 530~GHz for oblique incidence at 45$^\circ$, according to Eq.~\ref{eq:oblpath}. 
The latter option is selected for this work as it is more amenable to experimental characterization. 

\begin{figure}[tbp]
\centering
\includegraphics{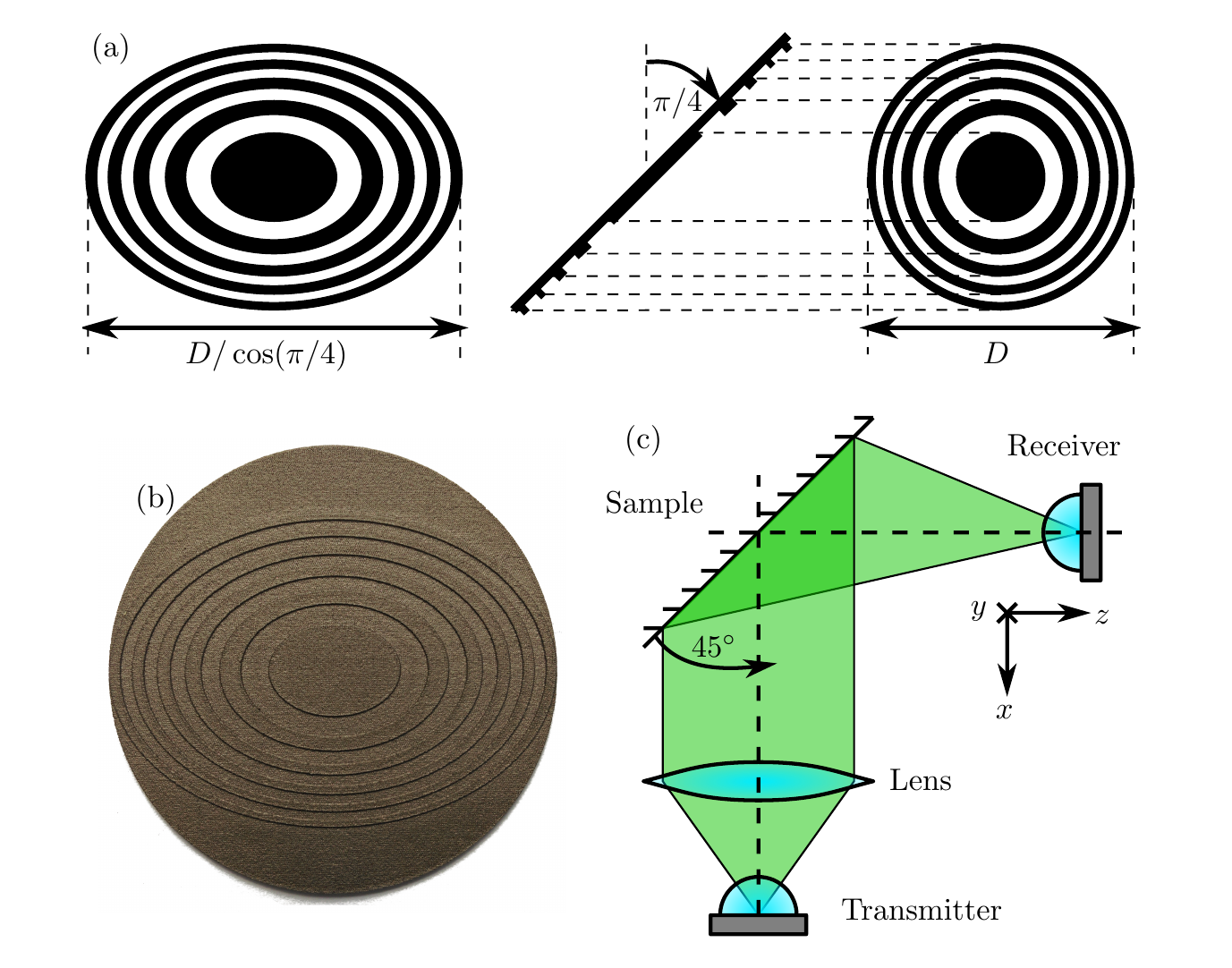}
	%
	%
	%
\caption{Zone plate design, showing (a) skewing of concentric circles into ellipses, (b) photograph of fabricated $\sim$50~mm-diameter sample, with chosen zone radii $r_m = 5.33$~mm, 7.54~mm, 9.25~mm, 10.70~mm, 11.98~mm, ..., and (c) measurement setup for oblique characterization of focal spot.
}
\label{fig:sample}
\end{figure}

A binary-phase zone plate is a diffractive optic that can both focus and collimate beams, in a manner that is analogous to a lens or parabolic reflector. 
This is achieved by imposing a phase profile with alternating rings phased at 0 and $\pi$ radians on the beam, with switching occurring at the radii dictated by the following expression, where $f$ is the focal length of the device \cite{attwood1999soft}, 
\begin{equation}
	r_m = \sqrt{ m \lambda f + \frac{m^2 \lambda^2}{4} }, \quad m=1,2,3,...
	\label{eq:zoneplate}
\end{equation}
Given that the reflective zone plate in the present work is operated at an oblique angle, elliptical rings are employed such that they present as circles when viewed at a 45$^\circ$ angle, as shown in Fig.~\ref{fig:sample}(a). 
The following equation describes the elliptical curve tracing the edge of zone $m$,
\begin{equation}
	\left( x \cos{\frac{\pi}{4}} \right) ^2  + y^2 = r_m^2.
	\label{eq:ellipse}
\end{equation}
A focal length of 50~mm is employed for this design, and the fabricated zone plate is shown in Fig.~\ref{fig:sample}(b).
As it is composed of solid metal, the realized device is significantly more robust than typical devices realized with microfabrication techniques. 

\subsection{Characterization and modeling of zone plate}

\begin{figure}[tbp]
\centering
\includegraphics{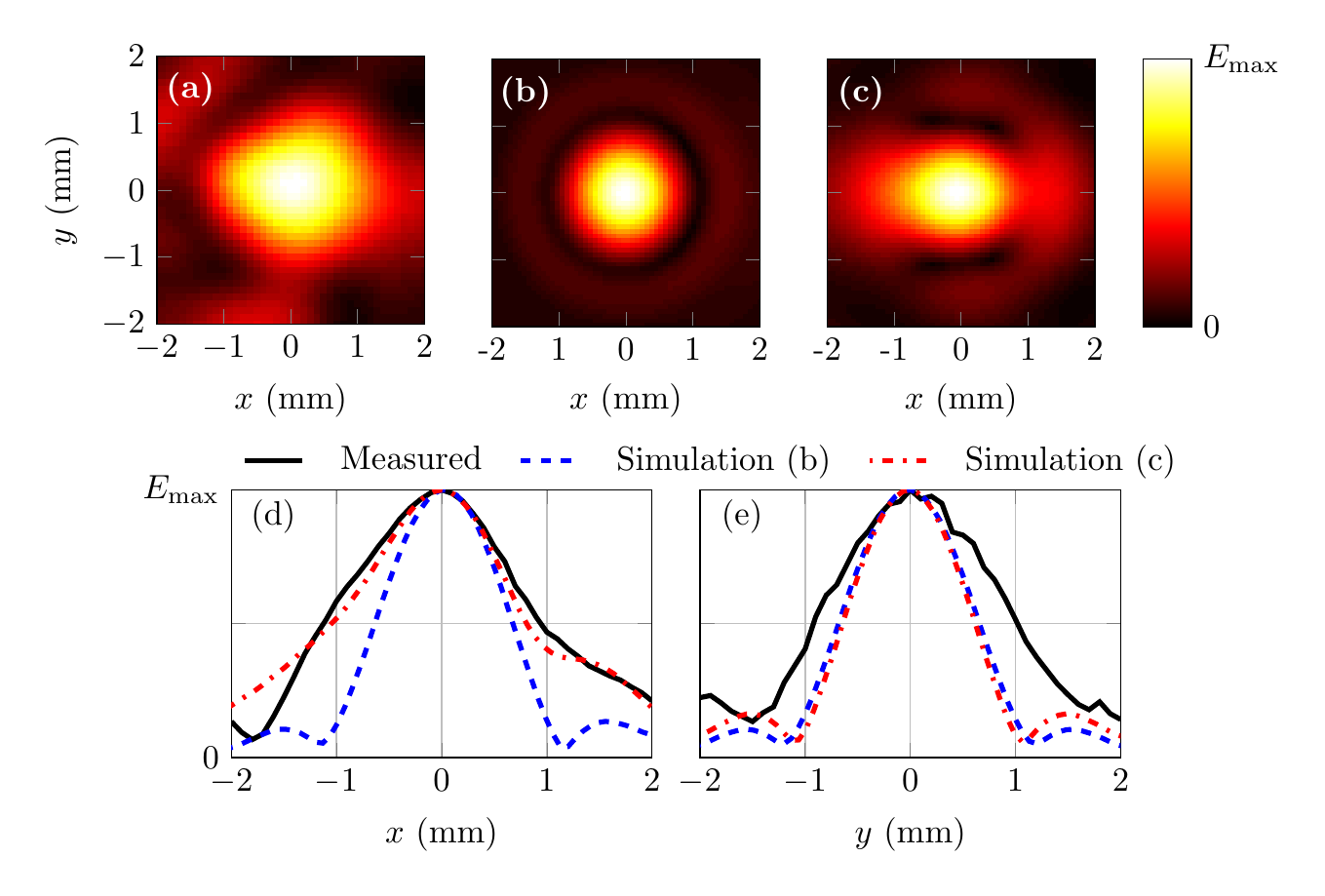}

	%

\caption{Field distribution in the focal plane, showing (a) measured linear amplitude distribution, and (b) simulated results, and (c) simulated result, incorporating 3$^\circ$ angular misalignment. 
For closer comparison, cross-sectional field profiles are given in (d) and (e).}
\label{fig:results}
\end{figure}

A fiber-coupled THz-TDS system (Menlo TERA K15) is employed to characterize the fabricated sample, with experimental setup shown in Fig.~\ref{fig:sample}(c). 
The sample is set at a 45$^\circ$ angle with respect to the transmitter, and is excited with a collimated beam. 
This beam is best approximated by a Gaussian beam, of $\sim$17~mm radial beamwidth, which is truncated to a 17~mm radius. 
TE-polarized light is employed, but this is not expected to impact the relative path length of the alternating zones, and hence it has no effect on diffractive behavior. 
The receiver raster-scans in the focal plane in order to image the focal spot. 
The measured amplitude profile is shown in Fig.~\ref{fig:results}(a), and a focal spot can be seen, albeit with some aberration and ringing effects. 
Additionally, the beam is slightly wider in the $x$-dimension than in the $y$-dimension. 
In order to evaluate the measured results, the  device is simulated using a procedure involving both full-wave simulations and and the Huygens-Fresnel principle \cite{goodman2005introduction}, with details given in Appendix~\ref{simulation}.
The result of this simulation is shown in Fig.~\ref{fig:results}(b), and a focal spot is clearly visible.
Furthermore, the beam widths in the $x$- and $y$-dimensions appear to be equal, and there is far less aberration than in the measured results. 

The most likely explanation for the disparity in $x$ and $y$ beam waist in the measured results is a slight rotational misalignment of the sample. 
This shortens the effective aperture of the zone plate in the x-dimension, which results in a broader focal spot. 
The rotational misalignment can be incorporated into the numerical simulation. 
An error of 3$^\circ$ results in the focal spot shown in Fig.~\ref{fig:results}(c).
It can be seen that this rotational misalignment has resulted in a focal spot that is broader in the $x$-dimension than in the $y$-dimension, much like the measured focal spot given in Fig.~\ref{fig:results}(a). 
Additionally, a possible explanation for general aberration in the focal spot is the variation in surface height over a more gradual scale than the characterized surface roughness.
This imparts some randomness to the phase of the reflected beam, which slightly degrades overall focal spot quality. 

In order to facilitate closer comparison, cross sectional field distributions of measured and simulated focal spots along the $x$- and $y$-axes are given in Figs.~\ref{fig:results}(d,e). 
It can be seen from these results that the simulation incorporating rotational misalignment is a better match to the measured results in the $x$-dimension than the simulation without rotational alignment. 
Additionally, whilst there is approximate agreement, both simulated field profiles are narrower than the measured profile in the $y$-dimension.
This is likely due to degradations in beam quality imposed by surface randomness, and vertical tilt of the sample may also be a contributing factor. 

\section{Conclusion}

We have evaluated the applicability of 3D-printed metal to terahertz technology. 
A 3D-printed titanium alloy is identified as potentially suitable for reflective optics in the terahertz range. 
This alloy, characterized by using THz-TDS, is shown to have efficiency of reflection above 80\% for frequencies below 900~GHz, and above 90\% below the nominal operating frequency of 530~GHz. 
A model incorporating surface roughness and variation in oxide layer thickness is employed to explain the reflection characteristics of the 3D-printed titanium alloy.  
Based on this model, we conclude that the most significant contributors to loss are surface roughness, and the presence of oxide at the surface. 
Despite its present use to describe the reflectivity of a specific titanium alloy, the developed analytical model is general, and can be applied to other metals. 

As a proof-of-concept, a terahertz zone plate with an operating frequency of 530~GHz is printed, and the focal spot is characterized with THz-TDS. 
The device is significantly more robust and durable than comparable devices realized with microfabrication techniques. 
Other such zone plates can be produced on-demand to arbitrary specification, as the additive manufacture process is rapid and readily customizable. 
Furthermore, other reflective devices, including free-form optics, may be produced in the same way to serve arbitrary beam-shaping requirements. 
Additionally, guiding structures such as hollow-tube waveguides and mode converters may also be printed directly in metal, at scales that are challenging for conventional machining and assembly. 
Therefore, our work opens opportunities for rapid prototyping of numerous diverse types of devices for the manipulation of terahertz radiation. 

More advanced 3D-printers can provide finer resolution, potentially supporting beam shaping applications towards 1~THz. 
Additionally, the conductivity of the alloy employed in this work is fairly low for a metal \cite{welsch1993materials}.
Other metals are available for 3D-printing, including other alloys of titanium, steel, and aluminum \cite{frazier2014metal}, which may have higher conductivity, and importantly, may oxidize less readily. 
For instance, printing in AlSi$_{10}$Mg has previously been demonstrated in other applications \cite{brandl2012additive}, and this material has electrical conductivity in the order of a hundred times that of grade-5 titanium  \cite{uliasz2012influence}.
Therefore, such alloys may exhibit higher reflectivity than the material presented in this work.  

\section*{Acknowledgments}

This work was performed in part at the OptoFab node of the Australian National Fabrication Facility utilizing Commonwealth and South Australian State Government funding.
D.A. acknowledges funding from the Australian Research Council (ARC), grant number FT120100351.
We wish to thank Luis Lima-Marques and Lijesh Thomas from the University of Adelaide, Institude for Photonics and Advanced Sensing for conducting the 3D printing and characterizing the surface profile, and Ken Neuber of The University of Adelaide, Adelaide Microscopy for overseeing EDX characterization.

\appendix

\section{Material properties}

\subsection{Bulk metal}
\label{materialdrude}

For grade-5 titanium, the electrical conductivity at room temperature is $\sim0.68$~MS/m \cite{welsch1993materials}.
A Drude model of this material exists in the literature \cite{wieting1976infrared}, indicating  plasma and collision frequencies of $1.99\times10^{16}$~rad/s, and $5.12\times10^{15}$~rad/s, respectively. 
These values are employed to model the material properties of the titanium alloy. 

\subsection{Oxide layer}
\label{materialoxide}

The oxide layer is approximated as pure titania of variable thickness, as explained in Section~\ref{reflection}. 
The properties of titania are known in the literature. 
The relative permittivity of titania is $\epsilon_r = 109.96$ \cite{scheller2009modelling}.
The absorption coefficient, in cm$^{-1}$, is empirically modeled with a quadratic expression,  $\alpha = 25.5\nu^2 - 5.5\nu + 4.2$, where $\nu$ is the terahertz frequency  \cite{ung2011high}.

\section{Zone plate simulation}

\label{simulation}

The entire zone plate structure is simulated using the numerical electromagnetics package CST Microwave Studio.
Oblique plane-wave incidence is employed, in order to approximate a TE-polarized collimated beam impinging upon the surface of the structure with a 45$^\circ$ angle of incidence. 
The structure is electrically large, and hence it is necessary to approximate the metal with PEC, to reduce the simulation complexity.
This is valid, as the properties of the metal and the surface at subwavelength scales are not expected to impact diffractive behavior, but rather will only result in some losses. 
The resulting field distribution is extracted into a text file, for further processing.
In order to isolate the scattered field from the total field, a second simulation is performed considering only the  bounding box, and the resulting field distribution is subtracted from the field distribution of the zone plate simulation. 

The scattered field from the zone plate structure is imported into Matlab. 
It is necessary to account for oblique excitation, in order to effectively view the structure at a 45$^\circ$ angle. 
To this end, a linear phase profile of $ x  2\pi \sin(\pi/4) / \lambda$ is imposed on the field distribution, and the $x$-axis is shortened by a factor of $\cos(\pi/4)$.
This mathematical transformation is the equivalent of the procedure described in Fig.~\ref{fig:sample}.

The software package employed, CST Microwave Studio, is most amenable to plane-wave excitation. 
However the collimated beam that is employed in the measurement is finite in extent, and the width of the beam impacts the resulting focal spot. 
In order to better approximate the collimated beam of the fiber-coupled THz-TDS system employed, a Gaussian beam profile, of radial beamwidth 17~mm, and truncated to a 17~mm radius, is imposed on the scattered field distribution. 
Finally, the Hugens-Fresnel principle \cite{goodman2005introduction} is employed to forwards-propagate the resulting field profile to the focal plane, over a distance of 50~mm. 

In order to determine the effect of rotational misalignment experienced during experimental characterization of the zone plate, further alterations are made to the scattered field profile prior to employing the Huygens-Fresnel principle. 
Rotational misalignment is approximated by employing an angle other than $\pi/4$ in the transformation described above, which compensates for the oblique excitation. 
This effectively results in `viewing' the sample at the wrong angle. 

\end{document}